\documentclass[useAMS,usenatbib]{mn2e}
\usepackage[dvips]{graphicx}
\usepackage{multirow}
\usepackage{array}

\title[An improved reaction scheme for oxygen and hydrocarbons]
  {An improved chemical scheme for the reactions of atomic oxygen and simple unsaturated hydrocarbons - implications for star-forming regions}
\author[A. Occhiogrosso et al]
  {Angela~Occhiogrosso,$^1$\thanks{E-mail: ao@star.ucl.ac.uk.}
  Serena Viti,$^1$ \& Nadia Balucani$^2$ \\
  $^1$Dept. of Physics and Astronomy, UCL, Gower Place, London WC1E6BT, UK \\
  $^2$Dipartimento di Chimica, Universit{\`a} degli Studi di Perugia, Via Elce di Sotto 8, Perugia, 06123, Italy}
\date{Released 2013 Xxxxx XX}

\pagerange{\pageref{firstpage}--\pageref{lastpage}} \pubyear{2013}

\def\LaTeX{L\kern-.36em\raise.3ex\hbox{a}\kern-.15em
    T\kern-.1667em\lower.7ex\hbox{E}\kern-.125emX}

\begin{document}

\maketitle

\begin{abstract}
Recent laboratory experiments have demonstrated that, even though contribution from other reaction channels cannot be neglected, unsaturated hydrocarbons easily break their multiple C-C bonds to form CO after their interactions with atomic oxygen. Here we present an upgraded chemical modelling including a revision of the reactions between oxygen atoms and small unsaturated hydrocarbons for different astrochemical environments. A first conclusion is that towards hot cores/corinos atomic oxygen easily degrades unsaturated hydrocarbons directly to CO or to its precursor species (such as HCCO or HCO) and destroys the double or triple bond of alkenes and alkynes. Therefore, environments rich in atomic oxygen at a relatively high temperature are not expected to be rich in large unsaturated hydrocarbons or polycyclic aromatic hydrocarbons. On the contrary, in O-poor and C-rich objects, hydrocarbon growth can occur to a large extent. In addition, new radical species, namely ketyl and vinoxy radicals, generated from other reaction channels can influence the abundances of hydrocarbons towards hot cores. We, therefore, suggest they should be included in the available databases.
Hydrocarbon column densities are calculated in the 10$^{13}$-10$^{15}$ cm$^{-2}$ range, in good agreement with their observed values, despite the small number of data currently published in the literature.

\end{abstract}

\begin{keywords}
 astrochemistry -- stars:formation -- ISM:abundances -- ISM:molecules.
\end{keywords}

\section{Introduction}
The interstellar medium (ISM) is extremely heterogeneous with a broad range of temperatures, densities and extinctions (Snow \& Bierbaum 2008). 
As a consequence a wide variety of chemical processes can occur. 
Oxygen is an important player in the ISM chemistry because it is the most abundant element after hydrogen and helium.
Interstellar oxygen can be found in different forms and compounds, such  
as neutral or ionised gas, in the water of icy mantles or even 
in refractory materials. Its gas-phase abundances in the diffuse medium are well known 
(Meyer et al. 1997), while the oxygen chemistry in denser environments seems to be poorly understood  
due to observational constraints (Jensen et al. 2005). 
On the other hand oxygen chemistry has been widely investigated towards extragalactic and galactic environments: in line with the fact that the C/O ratio determines the 
kind of reactions taking place in a specific region of ISM (Tsuji 1973), Ag\'{u}ndez $\&$ Cernicharo (2006), 
looking at the circumstellar shells of IRC +10216, 
observed that carbon is locked either as CO or other O-bearing species when C/O $<$ 1, 
while a higher concentration of reduced C-bearing species is present 
when C/O $>$ 1. The latter condition leads to an increase in the growth of 
unsaturated hydrocarbons. 
This has been extensively confirmed by modelling and observations and in particular Woods et al. (2012) found a 
large presence of unsaturated hydrocarbons, such as acetylene, by computing the physical conditions around carbon-rich AGB stars at sub-solar metallicities. 

Unsaturated hydrocarbons are organic molecules characterised by double or triple covalent bonds between adjacent carbon atoms; 
presently, there is no generally accepted explanation of their formation mechanisms but, as mentioned above, their abundances 
seem to be inversely correlated to the availability of atomic oxygen (Fortney 2012). 
Their most likely reactions start 
with the addition of atomic and radical species to their multiple bonds. In most cases, the elimination of a hydrogen atom or of a group follows, 
leading to more complex compounds (see, for instance, the cases of the reactions C + C$_2$H$_2$, C$_2$ + C$_2$H$_2$, CN + C$_2$H$_2$,
Kaiser 2000, Leonori et al. 2008a, Kaiser et al. 2003, Leonori et al. 2008b, Huang et al. 2000, Leonori et al. 2010).
Recent experiments in the gas-phase (Capozza et al. 2004; Fu et al. 2012a; Fu et al. 2012b; Leonori et al. 2012) 
suggest that the C-C bond breaking channel with the formation of CO is an important reaction channel for the reaction O + C$_2$H$_2$ and O + C$_2$H$_4$, while
it is by far the dominant loss path for the reaction O + CH$_2$CCH$_2$ (allene). 
The fact that the reactions of atomic oxygen with unsaturated hydrocarbons lead in one step to CO can have a 
pivotal role in estimating the CO abundances both in the gas-phase and on grain surfaces. 
Carbon monoxide is in turn a key species for the synthesis of organic complex molecules 
on grain surfaces since it can easily hydrogenate to methanol as experimentally demonstrated by Modica \& Palumbo (2010). Another important aspect 
regarding these small hydrocarbons is their loss paths to form CH$_{2}$. The latter molecular radical is indeed known 
to be a potential precursor of polycyclic aromatic hydrocarbons (PAHs) and many studies have been performed to stress several implications due 
to the presence of these aromatic hydrocarbons on the dust (Verstraete 2011; Salama 2008); 
on the other hand small unsaturated hydrocarbons are the building blocks of polyynes 
(organic compounds with alternating C-C single and triple bonds) and cumulenes 
(hydrocarbons with three or more consecutive C-C double bonds) that are also known as starting points for the synthesis of PAHs.
    
In the present study we focus on the reactions between O atoms and small unsaturated hydrocarbons, 
namely, acetylene, ethylene, methylacetylene and allene, under interstellar medium conditions. 
In particular, we propose a revised chemical scheme (compared to those included in the available databases) 
where our reaction rates are based on new experimental branching ratios (BRs) for the different reaction channels (Capozza et al. 
2004; Fu et al. 2012a; Fu et al. 2012b; Leonori et al. 2012). 
In order to assess the effects of the recent laboratory data on the hydrocarbon abundances across different astrophysical environments, we model four typical astrochemical regions and we compare our theoretical results with the observations.
The paper is organised as follows: Section 2 describes the laboratory technique used and the revised chemistry we inserted in our model; Section 3 focuses on the modelling of four different astrochemical environments; results are discussed in Section 4 and our conclusions are summarised in Section 5.

\section{The reactions of atomic oxygen with small unsaturated hydrocarbons}
The following gas-phase neutral-neutral reactions have been experimentally investigated  
at the University of Perugia by means of the crossed molecular beam (CMB) technique with mass spectrometric (MS) detection.
\begin{enumerate}
\item O + C$_{2}$H$_{4}$ $\rightarrow$ products

\item O + C$_{2}$H$_{2}$ $\rightarrow$ products    

\item O + CH$_{3}$CCH $\rightarrow$ products

\item O + CH$_{2}$CCH$_{2}$ $\rightarrow$ products

\end{enumerate} 
In particular, the reaction mechanisms, the nature of the primary products 
of the different competing channels and their branching ratios have been determined (Capozza et al. 2004; Fu et al. 2012a; Fu et al. 2012b; Leonori et al. 2012). 
From the analysis of the measured product angular and velocity distributions, the product BRs are evaluated
(for the detailed procedure see Balucani et al. 2006 and Casavecchia et al. 2009).
To derive the rate for each reaction channel, 
we consider the global rate constants for O + C$_{2}$H$_{2}$, O + C$_{2}$H$_{4}$, O + CH$_{3}$CCH and O + CH$_{2}$CCH$_{2}$ scaled by the experimental BRs (see Table 1 and Table 2). For reactions (i) and (ii), the rate coefficients and temperature dependence are those recommended by Baulch et al. (2005) in the range 200-2500 K and 220-2000 K for the two reactive systems, respectively. For reactions (iii) and (iv), we have considered the rate constants and temperature dependence recommended by Cvetanovic (1987) in the ranges 290 - 1300 K and 290 - 500 K, respectively. The temperatures of interest in interstellar objects are lower than those for which the temperature dependencies have been determined. Therefore, we have extrapolated the trends down to 10 K, but this is an approximation and important deviations are possible.

The oxygen chemistry with hydrocarbons based on the new laboratory data was inserted in the UCL\_CHEM chemical model. 
UCL\_CHEM simulates the formation of a prestellar core (Phase I) and its subsequent evolution once a star is born (Phase II). 
During Phase I gas-grain interactions occur due to the freeze-out of atoms and molecules on the grain surfaces when temperatures drop down to 10 K. 
UCL\_CHEM accounts for the chemistry occurring in the gas-phase as well as on the grain surfaces during the free-fall collapse of a diffuse cloud. 
Species can also sublimate due to both non-thermal (10 K) and thermal desorption (200-300 K) (Phase II). 
A more detailed description of the model can be found in the literature (Occhiogrosso et al. 2011). 
Besides new experimental values for the reactions listed above, the new part of the chemical network contains a new species, ketyl radical (HCCO), which is produced in the reaction O + C$_{2}$H$_{2}$. 
To date, HCCO has not been detected in the interstellar medium, but its presence was already predicted in the past 
by Turner $\&$ Sears (1989) who provided an upper limit for the HCCO/H$_{2}$CCO ratio. 
We also introduce the vinoxy radical (CH$_{2}$CHO), produced in the reaction O + C$_{2}$H$_{4}$. 
Furthermore, in addition to methylacetylene, its isomer CH$_{2}$CCH$_{2}$ (allene or propadiene) is explicitly considered. 
We have used either laboratory data or the rate coefficients included in the KIDA database (Wakelam et al. 2012) 
for the reactions that involve these two hydrocarbon species. 
This distinction is not present to date in the UMIST database (Woodall et al. 2007)$^{1}$.
\footnotetext[1]{Just prior of submission of the present paper a new version of the UMIST database has been announced (McElroy et al. 2013) however UMIST 2012 database is not yet available.}

\begin{table*}
 \begin{minipage}{126mm}
  \caption{Rate constants (at 200 K, typical of a hot core) for the different competing reaction channels for ethylene and acetylene with oxygen atoms. The last column contains the formul$\ae$ (taken from NIST database) adopted in order to evaluate the rate constants in column 3.}
  \begin{tabular}{@{}lccc}
  \hline
Reaction & BRs/$\%$ & k/cm$^{3}$mol$^{-1}$s$^{-1}$ & k(T)$_{let}$/cm$^{3}$mol$^{-1}$s$^{-1}$\\
   \hline
   \hline
O + C$_{2}$H$_{2}$ $\rightarrow$ HCCO + H & 79 & 9.9$\times$10$^{-15}$ & \multirow{3}{*}{1.89$\times$10$^{-12}$ (T/298)$^{2.10}$ e$^{-6535/RT}$}\\
   \cline{1-3}
O + C$_{2}$H$_{2}$ $\rightarrow$ CH$_{2}$ + CO & 21 & 2.6$\times$10$^{-15}$ & \\
   \cline{1-3}
O + C$_{2}$H$_{2}$ $\rightarrow$ C$_{2}$H + OH & 0 & 0 & \\
   \hline
   \hline
O + C$_{2}$H$_{4}$ $\rightarrow$ CH$_{3}$ + HCO & 34 & 1.0$\times$10$^{-13}$ & \multirow{5}{*}{1.01$\times$10$^{-12}$(T/298)$^{1.88}$ e$^{-765/RT}$}\\ 
   \cline{1-3}   
O + C$_{2}$H$_{4}$ $\rightarrow$ CH$_{2}$CHO + H & 30 & 9.0$\times$10$^{-14}$ & \\
   \cline{1-3}  
O + C$_{2}$H$_{4}$ $\rightarrow$ H$_{2}$CO + CH$_{2}$ & 20 & 6.0$\times$10$^{-14}$ &\\
   \cline{1-3}  
O + C$_{2}$H$_{4}$ $\rightarrow$ CH$_{2}$CO + H$_{2}$ & 13 & 3.9$\times$10$^{-14}$ & \\
   \cline{1-3}   
O + C$_{2}$H$_{4}$ $\rightarrow$ CH$_{3}$CO + H & 3 & 9.0$\times$10$^{-15}$ & \\
   \cline{1-3}
O + C$_{2}$H$_{4}$ $\rightarrow$ C$_{2}$H$_{3}$ + OH & 0 & 0 & \\
   \hline
  \end{tabular}
 \end{minipage}
\end{table*}

\begin{table*}
 \begin{minipage}{126mm}
  \caption{Rate constants (at 200 K, typical of a hot core) for the different competing reaction channels. The last column contains the formul$\ae$ (taken from NIST database) adopted in order to evaluate the rate constants in column 3.}
  \begin{tabular}{@{}lccc}
   \hline
Reaction & BRs/$\%$ & k/cm$^{3}$mol$^{-1}$s$^{-1}$ & k(T)$_{let}$/cm$^{3}$mol$^{-1}$s$^{-1}$\\
   \hline
   \hline
O + CH$_{3}$CCH $\rightarrow$ C$_{2}$H$_{4}$ + CO & 68.0 & 5.2$\times$10$^{-13}$ & \multirow{5}{*}{2.18$\times$10$^{-11}$ e$^{-831/RT}$}\\
   \cline{1-3}
O + CH$_{3}$CCH $\rightarrow$ C$_{2}$H$_{3}$ + HCO & 10.0 & 7.7$\times$10$^{-14}$ & \\
   \cline{1-3}
O + CH$_{3}$CCH $\rightarrow$ C$_{2}$H$_{2}$ + H$_{2}$CO & 10.0 & 7.7$\times$10$^{-14}$ & \\
   \cline{1-3}
O + CH$_{3}$CCH $\rightarrow$ CH$_{3}$CCO + H & 5.5 & 4.3$\times$10$^{-14}$ & \\
   \cline{1-3}
O + CH$_{3}$CCH $\rightarrow$ CH$_{3}$ + HCCO & 0.4 & 3.1$\times$10$^{-15}$ & \\
   \hline
   \hline
O + CH$_{2}$CCH$_{2}$ $\rightarrow$ C$_{2}$H$_{4}$ + CO & 81.5 & 9.8$\times$10$^{-13}$ & \multirow{5}{*}{2.82$\times$10$^{-11}$ e$^{-7732/RT}$}\\
   \cline{1-3}
O + CH$_{2}$CCH$_{2}$ $\rightarrow$ C$_{2}$H$_{2}$ + H$_{2}$CO & 9.6 & 1.1$\times$10$^{-13}$ & \\
   \cline{1-3}    
O + CH$_{2}$CCH$_{2}$ $\rightarrow$ C$_{2}$H$_{3}$ + HCO & 7.0 & 8.4$\times$10$^{-14}$ & \\
   \cline{1-3}
O + CH$_{2}$CCH$_{2}$ $\rightarrow$ CH$_{2}$CCHO + H & 1.6 & 1.9$\times$10$^{-14}$ & \\
   \cline{1-3}
O + CH$_{2}$CCH$_{2}$ $\rightarrow$ CH$_{2}$CO + CH$_{2}$ & 0.3 & 3.6$\times$10$^{-15}$ & \\
   \hline
  \end{tabular}
 \end{minipage}
\end{table*}

We have investigated how the revised part of the chemical model affects the trends in the hydrocarbon fractional 
abundances (relative to hydrogen nuclei) at typical densities in the ISM (Figure 1). 
Then, we have run a grid of models representing (i) a diffuse cloud, (ii) a translucent cloud, (iii) 
a dark core and (iv) a hot core. 
We will analyse the contribution to the hydrocarbon abundances from their primary products 
formed during their interactions with oxygen atoms. We will 
finally compare our theoretical hydrocarbon abundances with those from 
the observations in cold and warm interstellar medium environments, where these species have been detected.

\subsection{The reactions O + C$_{2}$H$_{2}$ and O + C$_{2}$H$_{4}$}
Ethylene or ethene, C$_{2}$H$_{4}$, is the simplest 
alkene (alkenes are hydrocarbons with a carbon-carbon double bond), while acetylene or ethyne, 
C$_{2}$H$_{2}$, is the smallest alkyne (alkynes are hydrocarbons with a carbon-carbon triple bond). 
These two molecules have a similar structure,
but their reactivity with oxygen atoms seems to be different (see Table 1). 
In particular, because of the larger number of atoms, ethylene leads to the formation of a greater number of products. 
According to experimental results (Capozza et al. 2004, Fu et al. 2012a,b)
in both cases the channels leading to OH formation are inefficient; this was accounted for
in the UMIST 2006 database (Woodall et al. 2007). 
The main channels of the reaction O + ethylene have been found 
to be those leading to CH$_{3}$ + HCO (34\%) and CH$_{2}$CHO + H (30\%) (Fu et al. 2012a,b). 
Regarding the reaction between acetylene and atomic oxygen, 
the dominant channel is the one leading to HCCO + H with a BR of 79\% while the channel leading to CH$_{2}$ and CO accounts for the remaining 21\% (Capozza et al. 2004; Leonori et al. 2013).
We find that HCCO plays a pivotal role in the trends 
of C$_{2}$H$_{2}$ abundances (see Section 3 for more details). 
To date the ketyl radical has not been included in any database. We have also introduced a chemistry network for the loss processes of HCCO as shown in Table 3. 

\begin{table}
 \begin{minipage}{126mm}
  \caption{Loss processes for HCCO}
  \label{anymode}
  \begin{tabular}{lc}
  \hline
Reaction & k/cm$^{3}$mol$^{-1}$s$^{-1}$ at 300 K\\
  \hline
$^{a}$H + HCCO $\rightarrow$ CO + CH$_{2}$  & 1.7$\times$10$^{-10}$\\

$^{a}${\bf H$_{2}$ + HCCO $\rightarrow$ products}  & 2.8$\times$10$^{-14}$\\ 

$^{a}$O + HCCO $\rightarrow$ CO + CO + H  & 1.6$\times$10$^{-10}$\\
  \hline
  \end{tabular}
\footnotetext[1]{Baulch et al. (2005) and references therein}
 \end{minipage}
\end{table}

\subsection{The reactions O + CH$_{3}$CCH (methylacetylene) and O + CH$_{2}$CCH$_{2}$ (allene)}

Since the experimental results by Leonori et al. (2012) and Balucani et al. (in preparation) have demonstrated that the reactions O + CH$_{3}$CCH and O + CH$_{2}$CCH$_{2}$ are characterised by different  branching ratios (and also their rate coefficients are different), we have separated the two isomeric species in our codes. The UMIST 2006 database (Woodall et al. 2007) does not distinguish between the two C$_{3}$H$_{4}$ isomers. Therefore, it is implicitly assumed that
CH$_{3}$CCH and CH$_{2}$CCH$_{2}$ are formed and destroyed in the same reactions and with the same efficiency. Such an assumption, however, is not warranted.
For instance, it has been suggested that an important contribution to the formation
 of both isomers is the reaction between methylidyne radical (CH) and ethylene (Gosavi et al. 1985;
 Wang et al. 1998). Nevertheless, recent laboratory investigations (Zhang et al. 2011) have shown 
allene to the main product with a ratio of 75-80\%.
To account separately for CH$_{3}$CCH and CH$_{2}$CCH$_{2}$ formation we have inserted the main path rate coefficients as recommended by KIDA database (Wakelam et al. 2012) (see Table 4). 
In Table 2 the open channels for reactions (iii) and (iv) are reported with their relative yields, as experimentally determined (Leonori et al. 2012; Balucani et al., in preparation)
We want to point out 
that, at this stage, we do not distinguish between the two C$_{3}$H$_{3}$O isomers, CH$_{3}$CCO 
(for the case of methylacetylene) and CH$_{2}$CCHO (for the case of allene); however we plan to 
investigate their chemistry in a future work.
In Table 2, the main dissimilarity is that, in the case of allene, the channel for the formation of 
CO dominates with a ratio of 81.5\%, while 
the BR of the reaction producing carbon monoxide from methylacetylene is only 68\% 
(Balucani et al., in preparation). 
The other channels contribute to a minor extent, especially in the case of the O + 
CH$_{2}$CCH$_{2}$ reaction.

\begin{table*}
 \begin{minipage}{126mm}
  \caption{Paths of formation for C$_{3}$H$_{5}$, CH$_{3}$CCH and CH$_{2}$CCH$_{2}$ based on data from KIDA database (Wakelam et al. 2012).}
  \label{anymode}
  \begin{tabular}{@{}lccc}
   \hline
Reaction & $\alpha$ & $\beta$ & $\gamma$ \\
   \hline
H + CH$_{3}$CHCH$_{2}$ $\rightarrow$ C$_{3}$H$_{5}$ + H$_{2}$ & 4.47$\times$10$^{-13}$ & 2.50 & 1250\\
C$_{2}$H$_{5}$ + CH$_{3}$CHCH$_{2}$ $\rightarrow$ C$_{3}$H$_{5}$ + C$_{2}$H$_{6}$ & 2.01$\times$10$^{-11}$ & 0 & 0\\
C$_{2}$H$_{3}$ + CH$_{3}$CHCH$_{2}$ $\rightarrow$ C$_{3}$H$_{5}$ + C$_{2}$H$_{4}$ & 1.72$\times$10$^{-15}$ & 3.50 & 2360\\
CH$_{2}$ + CH$_{3}$CHCH$_{2}$ $\rightarrow$ C$_{3}$H$_{5}$ + CH$_{3}$ & 2.70$\times$10$^{-12}$ & 0 & 2660\\
CH$_{3}$ + C$_{2}$H$_{3}$ $\rightarrow$ C$_{3}$H$_{5}$ + H & 1.20$\times$10$^{-10}$ & 0 & 0\\
CCH + CH$_{3}$CHCH$_{2}$ $\rightarrow$ C$_{3}$H$_{5}$ + C$_{2}$H$_{2}$ & 2.40$\times$10$^{-10}$ & 0 & 0\\
   \hline
C$_{3}$H$_{5}$$^{+}$ + NH$_{3}$ $\rightarrow$ CH$_{3}$CCH + NH$_{4}$$^{+}$ & 9.00$\times$10$^{-10}$ & 0 & 0\\
C$_{2}$H$_{6}$ + CH$_{2}$CCH $\rightarrow$ CH$_{3}$CCH + C$_{2}$H$_{5}$ & 5.83$\times$10$^{-14}$ & 3.30 & 9990\\
CH$_{4}$ + CH$_{2}$CCH $\rightarrow$ CH$_{3}$CCH + CH$_{3}$ & 1.74$\times$10$^{-14}$ & 3.40 & 11700\\
H$_{2}$ + CH$_{2}$CCH $\rightarrow$ CH$_{3}$CCH + H & 1.42$\times$10$^{-13}$ & 2.38 & 9560\\
C$_{2}$H$_{6}$ + CH$_{2}$CCH $\rightarrow$ CH$_{3}$CCH + C$_{2}$H$_{5}$ & 4.30$\times$10$^{-12}$ & 0 & -66\\
C$_{2}$H$_{3}$ + C$_{3}$H$_{5}$ $\rightarrow$ CH$_{3}$CCH + C$_{2}$H$_{4}$ & 4.00$\times$10$^{-12}$ & 0 & 0\\
CH$_{3}$ + C$_{3}$H$_{5}$ $\rightarrow$ CH$_{3}$CCH + CH$_{4}$ & 4.03$\times$10$^{-13}$ & 0.32 & -66\\
CCH + CH$_{3}$CHCH$_{2}$ $\rightarrow$ CH$_{3}$CCH + C$_{2}$H$_{3}$ & 2.00$\times$10$^{-11}$ & 0 & 0\\
OH + C$_{3}$H$_{5}$ $\rightarrow$ CH$_{3}$CCH + H$_{2}$O & 1.00$\times$10$^{-11}$ & 0 & 0\\
H + C$_{3}$H$_{5}$ $\rightarrow$ CH$_{3}$CCH + H$_{2}$ & 3.00$\times$10$^{-10}$ & 0 & 0\\
$^{a}$CH + C$_{2}$H$_{4}$ $\rightarrow$ CH$_{3}$CCH + H & 7.80$\times$10$^{-11}$ & 0 & 0\\
   \hline
C$_{3}$H$_{5}$ + C$_{3}$H$_{5}$ $\rightarrow$ CH$_{2}$CCH$_{2}$ + CH$_{3}$CHCH$_{2}$ & 1.40$\times$10$^{-13}$ & 0 & -132\\
C$_{2}$H$_{6}$ + CH$_{2}$CCH $\rightarrow$ CH$_{2}$CCH$_{2}$ + C$_{2}$H$_{5}$ & 5.83$\times$10$^{-14}$ & 3.30 & 9990\\
CH$_{4}$ + CH$_{2}$CCH $\rightarrow$ CH$_{2}$CCH$_{2}$ + CH$_{3}$ & 1.74$\times$10$^{-14}$ & 3.40 & 11700\\
H$_{2}$ + CH$_{2}$CCH $\rightarrow$ CH$_{2}$CCH$_{2}$ + H & 1.42$\times$10$^{-13}$ & 2.38 & 9560\\
C$_{2}$H$_{6}$ + CH$_{2}$CCH $\rightarrow$ CH$_{2}$CCH$_{2}$ + C$_{2}$H$_{5}$ & 1.60$\times$10$^{-12}$ & 0 & -66\\
C$_{2}$H$_{3}$ + C$_{3}$H$_{5}$ $\rightarrow$ CH$_{2}$CCH$_{2}$ + C$_{2}$H$_{4}$ & 2.00$\times$10$^{-12}$ & 0 & 0\\
CH$_{3}$ + C$_{3}$H$_{5}$ $\rightarrow$ CH$_{2}$CCH$_{2}$ + CH$_{4}$ & 4.03$\times$10$^{-13}$ & 0.32 & -66\\
H + C$_{3}$H$_{5}$ $\rightarrow$ CH$_{2}$CCH$_{2}$ + H$_{2}$ & 3.30$\times$10$^{-10}$ & 0 & 0\\
$^{a}$CH + C$_{2}$H$_{4}$ $\rightarrow$ CH$_{2}$CCH$_{2}$ + H & 3.12$\times$10$^{-10}$ & 0 & 0\\
   \hline 
  \end{tabular}
\footnotetext[1]{The $\alpha$ value is evaluated by the product between the ratio for each channel of reaction (Zhang et al. 2011) and the total rate constant of 3.9x10$^{-10}$ cm$^{3}$mol$^{-1}$s$^{-1}$}
 \end{minipage}
\end{table*}

\section{Chemical modelling}
We investigated how the new reaction networks included in the model affect the hydrocarbon 
abundances by running simple models at typical densities for the ISM in the 10$^{2}$-10$^{6}$ 
cm$^{-3}$ range; in particular, we simulate the case of an unphysical cloud at 10 K and constant 
density where we do not account for any freeze-out reaction during the collapse phase. We plot the 
fractional abundances obtained as outputs from these test models for C$_{2}$H$_{2}$, 
C$_{2}$H$_{4}$ and the two C$_{3}$H$_{4}$ isomers as a function of the density of the cloud at 
5$\times$10$^{5}$ yrs (see Figure 1). In order to highlight the changes in the hydrocarbon 
abundances, we run models with the same physical parameters as described above but including 
the same chemistry as in the UMIST 2006 database. In other words, the difference observed are only due to the new reaction scheme for reactions (i)-(iv) with all the other parameters kept constant.
\begin{figure*}
 \includegraphics[angle=270, width=170mm]{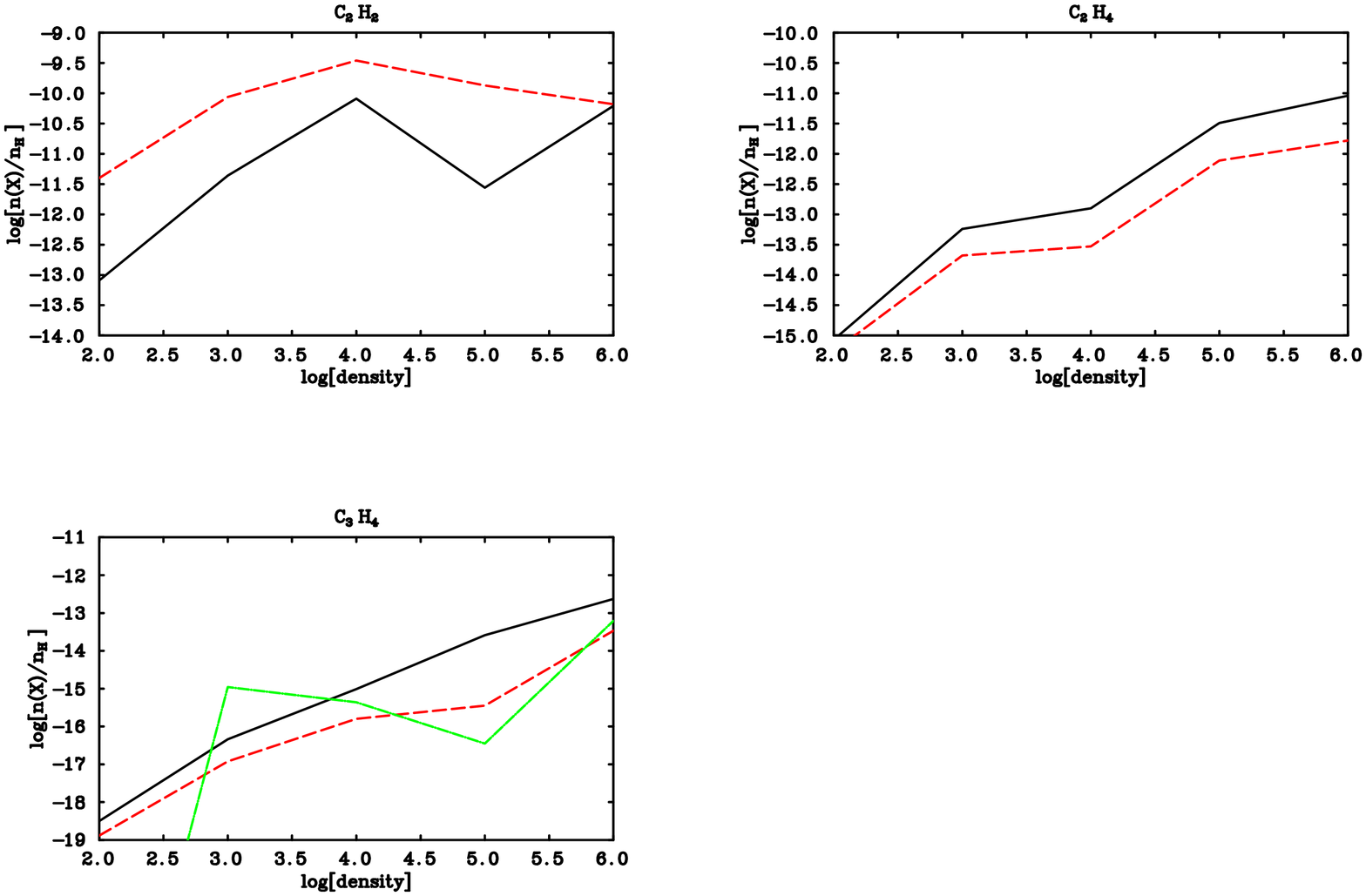}
 \caption{The fractional abundances (with respect to the total number of hydrogen nuclei) for selected species as a function of the density. In the Top panels the solid lines represent the outputs with the same chemistry as in UMIST 2006 database and the dashed lines are the outputs after the updating; Bottom panel shows the trends obtained from the revised chemistry suggested in the present study for CH$_{3}$CCH (dashed line) and CH$_{2}$CCH$_{2}$ (dotted line), while the solid line is the trend for both C$_{3}$H$_{4}$ isomers resulted from the previous version of the model.}
\end{figure*}
Figure 1 shows a comparison between the hydrocarbon fractional abundances (with respect to the 
total number of hydrogen atoms) from the original version of the code (solid line) and the ones 
from the models containing the revised chemistry (dashed line). In the case of C$_{3}$H$_{4}$ 
the solid line refers to the output obtained from the UMIST 2006 database that does not 
distinguish between the two isomers, while the dashed line and the dotted line are the trends for 
CH$_{3}$CCH and CH$_{2}$CCH$_{2}$, respectively, produced from the updated chemistry.
In the absence of freeze-out, for each selected density value we observe a discrepancy between 
the hydrocarbon fractional abundances before and after the updating, with the exception of 
C$_{2}$H$_{2}$ at 10$^{6}$ cm$^{-3}$. 
We emphasize that with this kind of comparison we are only analysing the effect of the new schemes on the hydrocarbon abundance prediction. We wish to stress that reactions (i)-(iv) are characterised by a certain activation energy and, therefore, their rate coefficients are very small at temperatures between 10-200 K. Nevertheless, they make a significant contribution in controlling the abundances of unsaturated hydrocarbons probably because of the large abundance of atomic oxygen. 

In order to understand the real influence of oxygen in the hydrocarbon chemistry, we modelled four 
different astronomical regions: a diffuse cloud, a translucent cloud, a dark cloud and a hot core. 
The following section describes the models in more details.

\subsection{Modelling of different astrochemical environments}
Table 5 lists the physical parameters used for our modelling. For each astronomical source 
we run two models: we employ a rate file with the same chemistry as in the UMIST 2006 database and 
another one with the revised chemical network proposed in the present study.
\begin{table*}
 \begin{minipage}{126mm}
  \caption{List of physical parameters for the case of a hot-core model: visual extinction (A$_{v}$), density (n$_{H}$), gas temperature (T) during Phase II, efficiency of the freeze-out (fr) during Phase I (the percentage of mantle CO (mCO) given by the freeze-out parameter by the end of Phase I of the chemical model).}
  \label{symbol}
  \begin{tabular}{@{}lcccc}
  \hline
 & A$_{v}$/mag & n$_{H}$/cm$^{-3}$ & T/K & fr/\%\\
   \hline
1. Diffuse cloud & 0-1 & 1$\times$10$^{2}$ & 100 & 0\\
2. Translucent Cloud & 1-5 & 1$\times$10$^{3}$ & 20 & 0\\ 
3. Dark core & 5-10 & 1$\times$10$^{4}$ & 10 & $\sim$40\\
4. Hot core & $\sim$500 & 1$\times$10$^{7}$ & 300 & $\sim$98\\
   \hline
  \end{tabular}
 \end{minipage}
\end{table*}
We start by considering the case for a diffuse cloud (Model 1), a quite extended low density (10$^{2}$ cm$^{-3}$) region permeated by the ultraviolet interstellar radiation field; temperatures go up to 100 K and the visual extinction is around 1 mag. Because of their relatively low-visual extinction ( A$_{v}$ $\sim$ 1 mag) diffuse clouds are dominated by photodissociation reactions and there is no chemistry occurring on the grain surfaces. 

As for the case of a diffuse cloud, atoms and molecules do not freeze-out on the dust for a translucent cloud (Model 2) either, an intermediate phase (in terms of physical conditions) between the diffuse and the dense medium (van Dishoeck 2000); the fact that they are called $translucent$ is due to the presence of a bright star in their proximity that allows absorption lines to be observed. Translucent clouds are relatively cold regions (20-50 K) where densities can reach values up to 10$^{3}$ cm$^{-3}$ although there is no evidence of a collapsing core inside these objects. Since translucent clouds can be penetrated by the radiation (their A$_{v}$ is between 1 and 5 mag), their chemistry is also affected by photoprocesses. Diffuse and translucent clouds are two examples of low-density interstellar regions.

We also explore the case for high-density sources, such as dark and hot cores, where a grain-surface chemistry is included in order to simulate the isothermal collapse phase during which atoms and molecules freeze onto the surfaces of dust and new species can be produced. This process is more efficient during the formation of a hot core (a very compact object characterised by densities $\ge$ 10$^{6}$ cm$^{-3}$) (Model 4) than for the case of a dark core (Model 3) where densities increase only up to 10$^{4}$ cm$^{-3}$; moreover while for the case of a hot core we need to model a second phase that describes all the heating effects that could induce molecules to thermally desorb (indeed temperatures rise up to 200 - 300 K), for a dark core we only account for the non-thermal desorption of species (for more details see Roberts et al. 2007) efficient already at 10 K.

\section{Results and Discussion}
We discuss our results only for the case of a hot core (Model 4). Results obtained as outputs from models 1, 2, 3 show some differences in the abundances of the species involved. Nevertheless we are not discussing those differences here since the resulting abundances are in all cases below the observational thresholds.
Concerning the hot core model, in order to simulate the warm-up of icy mantles and the consequent sublimation of molecules from their surfaces, we allow a time-step desorption of species as described in Viti et al. (2004). 
Based on laboratory data obtained from temperature programmed desorption experiments, Viti et al. (2004) categorised a list of species along with 
their desorption profiles in various temperature bands. C$_{2}$H$_{2}$ and  C$_{2}$H$_{4}$ were classified as intermediate between CO-like and H$_{2}$O-like species. 
Intermediate molecules were found in the hydrogenated ice layer although they were also able to escape via monomolecular desorption because of 
their moderate interaction with water ice. Since the two C$_{3}$H$_{4}$ isomers are non-polar organic species (as well as C$_{2}$H$_{2}$ and C$_{2}$H$_{4}$), 
we assumed for them a similar behaviour to the one for the case of acetylene and ethylene; we therefore classified methylacetylene and allene as intermediate. 

We analyse the outputs for the hydrocarbon abundances as well as the trends for the abundances of 
other species involved in the updated oxygen chemistry (see Figure 2.).

\begin{figure*}
 \includegraphics[angle=270, width=170mm]{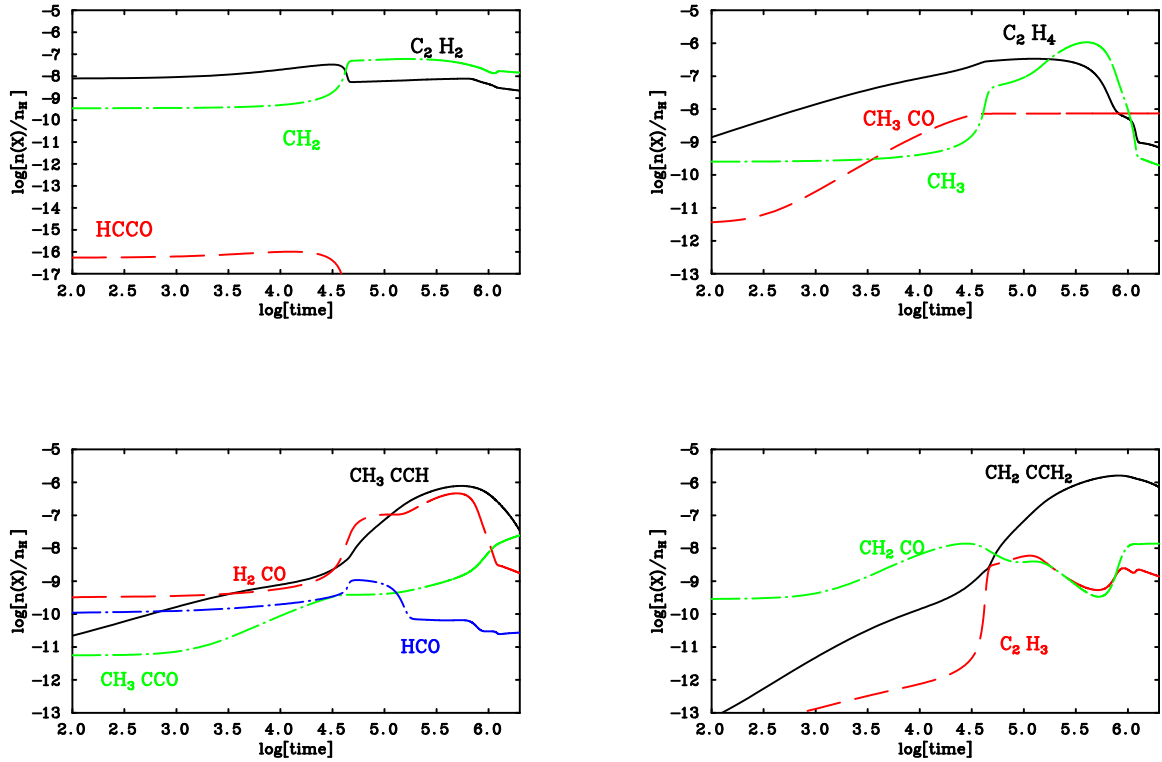}
 \caption{The fractional abundances (with respect to the total number of hydrogen nuclei) for selected species as a function of time.}
\end{figure*}
In particular, we focus on small patterns such as CH$_{2}$, CH$_{3}$, HCO, H$_{2}$CO that have been detected as products for all the hydrocarbon 
species considered. While it is difficult to determine the contribution of each reaction channel to the abundances of all hydrocarbons, UCL\_CHEM provides us with an analysis of the percentage of destruction and formation pathways for any selected species and we can therefore draw some conclusions on the trends displayed in Figure 2. 
For instance, CH$_{2}$ is known as one of the reactants in the synthesis of hydrocarbons and 
its abundance is expected to decrease when hydrocarbons form; however we observe a rise in the 
CH$_{2}$ abundances where the trends for CH$_{3}$CCH and CH$_{2}$CCH$_{2}$ show a peak; 
this behaviour could be explained in the light of the fact that HCCO drops to produce CH$_{2}$ 
and this reaction is five orders of magnitude more efficient than the reaction between O and acetylene or methylacetylene to form ketyl radical. Another important observation concerns the decrease in the ethylene abundances in order to produce the two C$_{3}$H$_{4}$ isomers, meaning that the reverse reaction is more likely than the channel for the formation of CO. CH$_{2}$CO is also produced by the reaction between O atoms and ethylene and as a consequence its abundances rise when the C$_{2}$H$_{4}$ profile drops just after 10$^{6}$ yrs. The major contribution to the formation for CH$_{3}$CCO is made by methylacetylene and, as a result, as soon as the latter species decreases, CH$_{3}$CCO increases in abundance. 
Concerning all the other molecules, it is difficult to make predictions since they can be formed by numerous reactions, 
but we estimate a high fractional abundance for C$_{2}$H$_{3}$ and H$_{2}$CO. This is in agreement with what we were expecting due to the 
fact that all hydrocarbons contribute to their formation. 
 
In order to validate the revised hydrocarbon chemistry, we evaluate the final column 
densities for most of the species involved and we compare our theoretical values with those 
from the observations towards hot cores. The column densities (in cm$^{-2}$) for a generic species 
Y are estimated by the approximation:
\begin{equation}
N(Y) \sim X(Y) \times A_{\rm v} \times N(H_{2}),
\end{equation} 
where X is the fractional abundance for the species Y, $A_{\rm v}$ is the visual extinction 
and $N(H_{2})$ is the column density of hydrogen at 1 mag and we set it equal to 1.6x10$^{21}$ 
cm$^{-2}$.
We also calculate the column densities obtained by the models without the new chemistry. 
Results are shown in Table 6. We observe that for all molecules, except for CH$_{3}$, the 
abundances obtained as outputs from the updated version of the code are one order of magnitude 
higher than those evaluated from its original version.

Hydrocarbons have been found across different astrochemical environments, from planet atmospheres 
to dark clouds, in the Milky way as well as towards other Galaxies (Pendleton 2004). One of the 
first hydrocarbon detections was performed by Ridgway et al. (1976), who observed CH$_{4}$ and 
C$_{2}$H$_{2}$ towards the supergiant IRC + 10216, where a few years later Betz (1981) found 
C$_{2}$H$_{4}$ with a column density of 10$^{16}$-10$^{17}$ cm$^{-2}$. 
Hydrocarbons were predicted to be abundant in the ISM but many of them were not observed at 
millimeter wavelengths because their symmetry forbids dipole rotational transitions. 
Some of the species listed in Table 6 have indeed not been detected towards hot cores yet, 
but since column densities for hydrocarbons are in the 10$^{13}$-10$^{15}$ cm$^{-2}$ range 
we expect the presence of similar amounts for C$_{2}$H$_{4}$ and CH$_{2}$CCH$_{2}$. 
A comparison between our theoretical results and data taken from the observations highlights a general agreement within an order of magnitude; in particular, after our updates, we achieve a greater match with the observed column densities of C$_{2}$H$_{2}$ and CH$_{2}$ than if we had used the chemistry in the UMIST 2006 database. Concerning the two C$_{3}$H$_{3}$O isomers, we only calculate an upper limit in their column densities since we do not account for a complete reaction network for their loss mechanisms. We would also like to stress that we have only qualitatively modelled a hot core without looking at a specific high-mass source therefore our theoretical column densities could be refined.
\begin{table*}
 \begin{minipage}{126mm}
  \caption{Comparison between observational and theoretical column densities (in cm$^{-2}$) for selected species. Starting from the left, the first column shows results from the original version of the code; on the second column we report the molecular column densities calculated after our updating.}
  \label{symbol}
  \begin{tabular}{@{}llllll}
  \hline
Molecule & UMIST 2006 & Present Work & Observations & Source & References\\
   \hline
C$_{2}$H$_{2}$ & 2.70$\times$10$^{15}$ & 4.08$\times$10$^{14}$ & 5.50$\times$10$^{14}$ & SgrA & Lacy et al. (1989)\\
C$_{2}$H$_{4}$ & 1.93$\times$10$^{13}$ & 1.26$\times$10$^{14}$ & - & - & -\\
CH$_{3}$CCH & 1.11$\times$10$^{15}$ & 6.18$\times$10$^{15}$ & 1.20$\times$10$^{15}$ & Orion-KL & Wang et al. (1993)\\
CH$_{2}$CCH$_{2}$ & not present & 1.78$\times$10$^{16}$ & - & - & -\\
CH$_{2}$ & 2.09$\times$10$^{15}$ & 4.88$\times$10$^{14}$ & 6.60$\times$10$^{13}$ & Orion-KL & Hollis et al. (1995)\\
CH$_{3}$ & 2.70$\times$10$^{15}$ & 7.73$\times$10$^{13}$ & 8.00$\times$10$^{14}$ & SgrA & Feuchtgruber et al. (2000) \\
HCCO & not present & 4.68$\times$10$^{3}$ & - & - & -\\
CH$_{2}$CO & 4.25$\times$10$^{14}$ & 2.55$\times$10$^{15}$ & 1.00$\times$10$^{14}$ & SgrB2 & Turner et al. (1977)\\
HCO & 5.62$\times$10$^{12}$ & 5.13$\times$10$^{12}$ & 1.03$\times$10$^{12}$ & W3 & Snyder et al. (1976)\\
H$_{2}$CO & 6.55$\times$10$^{14}$ & 1.85$\times$10$^{14}$ & 8.32$\times$10$^{14}$ & Orion-KL & Wright et al. (1996)\\
CH$_{3}$CO & not present & 9.13$\times$10$^{14}$ & - & - & -\\ 
CH$_{3}$CCO & not present & 5.80$\times$10$^{15}$ & - & - & -\\
C$_{2}$H$_{3}$ & 5.44$\times$10$^{13}$ & 1.30$\times$10$^{13}$ & - & - & -\\
   \hline
  \end{tabular}
 \end{minipage}
\end{table*}

\section{Conclusions}
In this paper we have presented how a more accurate chemical network representing the reactions 
of atomic oxygen with several unsaturated hydrocarbons affects the
abundances of simple organic species in ISM. We mostly focus on the modelling of hot ISM 
regions where these variations were found to be more evident; however colder regions such as 
diffuse and translucent clouds or a dark core have also been investigated. A first conclusion 
is that towards hot cores/corinos atomic oxygen easily degrades unsaturated hydrocarbons directly 
to CO or to its precursor species
(such as HCCO or HCO) and destroys the double or triple bond of alkenes and alkynes. 
Therefore, environments rich in atomic oxygen at a relatively high temperature are
not expected to be characterised by the presence of large unsaturated hydrocarbons or 
polycyclic aromatic hydrocarbons. On the contrary, in O-poor and C-rich objects,
hydrocarbon growth can occur to a large extent. In colder objects, the effect of the 
reactions of atomic oxygen with small unsaturated hydrocarbons is reduced
because of the presence of a small activation energy for these reactions. 
Moreover we highlight the main contribution from other reaction channels that are not yet 
included in the databases available: ketyl and vinoxy radicals are among the new significant 
species 
that should be considered in the modelling. New attempts to detect them should be pursued and a 
chemical scheme for the vinoxy radical loss pathways devised.
Furthermore, we have established that structural isomers, such as methylacetylene and allene, 
should be treated separately 
when they manifest a different chemical behaviour.
A more general conclusion is that, if we wish to fully understand the formation of the numerous 
organic species detected in ISM, the intricate chemistry that leads to their formation should 
be treated in detail.

We finally plan to refine our theoretical results by modelling specific sources where 
hydrocarbons have been found. In the present study we calculated the hydrocarbons column 
densities that match those from the observations within an order of magnitude although we 
were unable to make a comparison for all the species considered: most hydrocarbons are indeed 
non-polar molecules and therefore they cannot be detected. Despite this constraint, we aim to 
estimate the column densities for the case of symmetric hydrocarbons  based on the abundances of 
their derivative species; for instance, as already stated by Quan $\&$ Herbst (2007), the amount 
of allene in a source could be related to the abundances of its derivative cyanoallene that, 
contrary to allene, is detectable.

\section{Acknowledgments}
The research leading to these results has received funding from the [European Community's] Seventh Framework Program [FP7/2007-2013] 
under grant agreement n$^{\circ}$ 238258.
We acknowledge the COST Action CM0805 - The Chemical Cosmos: Understanding Chemistry in Astronomical Environments for supporting of a short-term mission of A. Occhiogrosso from UCL to Perugia.
The authors would also like to thank Paul Woods for useful discussions.

\end{document}